\documentclass[fleqn,10pt]{wlscirep}
\usepackage[utf8]{inputenc}
\usepackage[T1]{fontenc}
\usepackage{graphicx,color}
\usepackage{amsmath,amssymb,amsfonts,bm}

\title{Spin Hydrodynamic Generation in the Charged Subatomic Swirl}

\author[1]{Xingyu Guo}
\author[2,*]{Jinfeng Liao}
\author[1,**]{Enke Wang}
\affil[1]{Guangdong Provincial Key Laboratory of Nuclear Science, Institute of Quantum Matter, South China Normal University, Guangzhou 510006, China.}
\affil[2]{Physics Department and Center for Exploration of Energy and Matter,
Indiana University, 2401 N Milo B. Sampson Lane, Bloomington, IN 47408, USA.}

\affil[*]{liaoji@indiana.edu}
\affil[**]{wangek@scnu.edu.cn}

\begin{abstract}
Recently there have been significant interests in the spin hydrodynamic generation phenomenon from multiple disciplines of physics. Such phenomenon arises from global polarization effect of microscopic spin by macroscopic fluid rotation and is expected to occur in the hot quark-gluon fluid (the ``subatomic swirl'') created in relativistic nuclear collisions. This was indeed discovered in experiments which however revealed an intriguing puzzle: a polarization difference between particles and anti-particles. We suggest a novel application of a general connection between rotation and magnetic field: a magnetic field naturally arises along the fluid vorticity in the charged subatomic swirl. We establish this mechanism as a new way for generating long-lived in-medium magnetic field in heavy ion collisions. Due to its novel feature,  this new magnetic field provides a nontrivial explanation to the puzzling observation of a difference in spin hydrodynamic generation for particles and anti-particles in heavy ion collisions.  
\end{abstract}
\begin{document}

\flushbottom
\maketitle

\thispagestyle{empty}
\newpage

\section*{Introduction}

Recently there have been rapidly increasing interests in the understanding of properties and novel phenomena in many-body systems under the influence of extreme fields like strong magnetic field or fluid rotation. Such interests come across multiple disciplines like condensed matter physics, cold atomic gases, astrophysics and nuclear physics, see e.g.~\cite{spinhydro,Gooth:2017mbd,Fetter:2009zz,2008PhRvA..78a1601U,2009PhRvA..79e3621I,Berti,Watts:2016uzu,Grenier:2015pya,Yamamoto:2013zwa,Son:2012bg,Li:2014bha,2015PhRvX...5c1023H,2016NatCo...711615A,Basar:2013iaa,Miransky:2015ava,Fukushima:2018grm,Kharzeev:2015znc,Liao:2014ava,Hattori:2016emy}.  These extreme fields can induce nontrivial anomalous chiral transport effects such as the Chiral Magnetic Effect (CME)~\cite{Kharzeev:2007jp,Fukushima:2008xe,STAR_LPV1,STAR_LPV_BES} and Chiral Vortical Effect (CVE)~\cite{Son:2009tf,Kharzeev:2010gr,Landsteiner:2011iq} that have been enthusiastically studied. These extreme fields can also strongly influence the phase structures and phase transitions in various physical systems~\cite{Jiang:2016wvv,Ebihara:2016fwa,Chen:2015hfc,Chernodub:2017ref}.

In the context of heavy ion collisions, not only an extremely hot subatomic material known as a quark-gluon plasma (QGP) is created, there also exists the largest fluid vorticity  as well as the strongest magnetic fields~\cite{Fukushima:2018grm,Kharzeev:2015znc,Liao:2014ava}. These experiments are carried out at the Relativistic Heavy Ion Collider (RHIC) of BNL and the Large Hadron Collider (LHC) of CERN. Great efforts have been made to look for effects from these extreme fields, with intriguing evidences yet also with outstanding puzzles.

The fluid vorticity originates from the large angular momentum carried by the colliding system and has been quantitatively simulated with various tools~\cite{thermal_vorticity,Becattini:2014yxa,Becattini:2013vja,Csernai:2013bqa,Csernai:2014ywa,Becattini:2015ska,Jiang:2016woz,Shi:2017wpk,Becattini:2016gvu,Li:2017slc,Sun:2017xhx}. The large vorticity could  lead to observable effects such as global spin polarization of produced particles~\cite{Liang:2004ph,Gao:2007bc,Voloshin:2004ha,Betz:2007kg,Becattini:2007sr}.   Recently the STAR Collaboration at RHIC measured this effect for the hyperons and anti-hyperons~\cite{STAR:2017ckg}, extracting an average fluid vorticity of about $10^{21}\, sec^{-1}$. However, there is a visible difference in the polarization between hyperons and anti-hyperons, with a larger signal for the latter. Current data show a clearly nonzero mean value for their polarization difference especially in the $10\sim 20 \rm GeV$ beam energy region (corresponding ions in the beam flying at about $99\%$ of the speed of light), albeit with large error bars. Attempts were made to explain this puzzle but so far inconclusive~\cite{Becattini:2016gvu,Csernai:2018yok,Muller:2018ibh}. A sufficiently long-lived  magnetic field could provide such a  splitting but it is unclear how to generate that field.

The magnetic field in these collisions, though not strong enough to visibly influence the bulk medium collective dynamics, plays a central role for inducing the interesting effects such as the CME  in QGP.  
While the initial vacuum magnetic field (mainly from spectators) reaches a few times pion-mass-square (or $\sim 10^{14}$ Tesla), it lasts only for too short a time duration~\cite{Bloczynski:2012en,McLerran:2013hla,Gursoy:2014aka,Tuchin:2015oka,Inghirami:2016iru,Gursoy:2018yai}. A pressing puzzle here is whether certain mechanism could lead to a considerably long-lived in-medium magnetic field.

In this paper, we suggest a novel application of a general link  between rotation and magnetic field in  a charged fluid system to the swirling subatomic fluid created in heavy ion collisions and show how this helps provide resolutions to the puzzles discussed above. The link is that  a magnetic field naturally arises along the fluid vorticity direction from the currents associated with the swirling charges, a mechanism to be demonstrated in details later.   
While this basic mechanism is simple and generic, it has not been previously applied to a very distinctive fluid system --- the hot subatomic fluid consisting of strongly interacting elementary particles such as quarks, gluons as well as hadrons. This highly relativistic fluid is at extreme among various fluid systems ever achieved in laboratories, with the highest temperature ($\sim 10^{12}\rm K$), flowing over the smallest spatial scale ($\sim 10^{-15}\rm m$) and shortest time scale ($\sim 10^{-23}\rm sec$). The present study will take a novel step to expand the territory of the mechanism into such hitherto unexplored extreme regime and establish its presence in the charged subatomic swirl. By using information about the fluid vorticity and net electric charge density (particularly in low beam energy region) from nuclear stopping in heavy ion collisions, we will estimate the magnitude of this new magnetic field. Furthermore, we will show that a novel feature of such a magnetic field is its considerably long lifetime (as compared with any previously known source of magnetic field in heavy ion collisions), due to the persistence of fluid vorticity (by virtue of angular momentum conservation). This feature turns out to be crucial in making important contributions to the spin hydrodynamic generation in heavy ion collisions and providing a nontrivial explanation of the observed difference in particle/anti-particle global polarization.

\section*{Demonstration of the Mechanism}

The main purpose of this Section is to demonstrate the aforementioned  mechanism, i.e. the generation of magnetic field by swirling charges. This connection is to be explicitly shown both at single-particle level and at many-particle level in the fluid dynamics framework. In the last subsection, we derive a concrete relation to connect magnetic field and fluid vorticity in a charged fluid vortex model, which shall then be applied later for estimating magnetic field in heavy ion collisions. 

\subsection*{Magnetic field of a swirling  charged particle}

\begin{figure}[htb!]
\begin{center}
\includegraphics[width=10.cm]{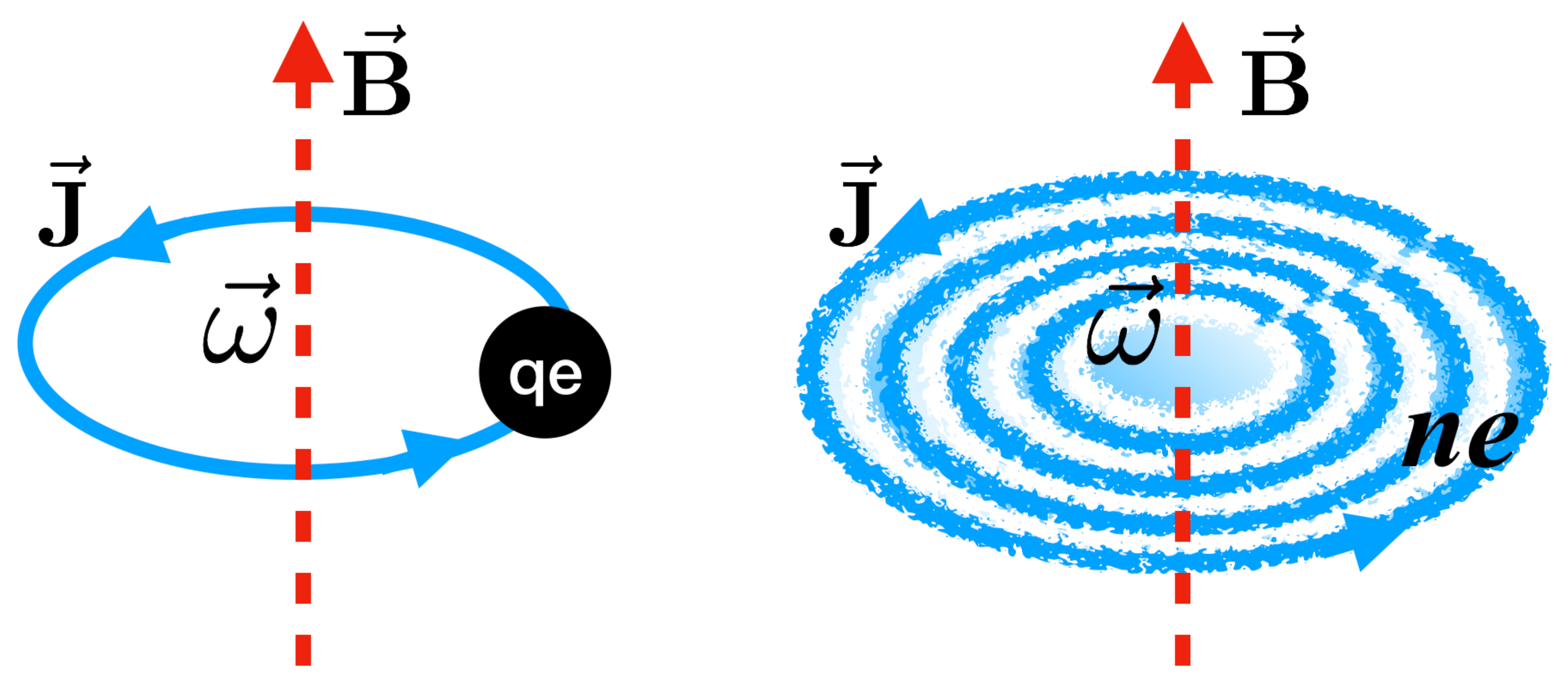}
\caption{\label{fig0} (color online) Illustration of the magnetic field generated by a single swirling charged particle (left) or by a swirling fluid with nonzero charge density (right).}
\end{center}
\end{figure}

We first demonstrate the main point, i.e. relation of magnetic field and rotation for a charged system, with the example of a single charged particle: see Fig.~\ref{fig0} (left).  

Let us start with  the simplest case, a classical relativistic charged particle (with charge $q e$ and mass $m$), undergoing a uniform circular motion at an angular speed $\omega_0$ with a radius $\rho_0$. The corresponding electric current is simply $I = \frac{qe\omega_0}{2\pi}$. Let us set up a cylindrical coordinate system $(\rho,\phi,z)$ with the circle on the $z=0$ plane and the center of the circle at the origin. The magnetic field within the circle on the $z=0$ plane points along the $\hat{z}$ direction and is given by: 
\begin{eqnarray}
B_z (\rho) &=& B_0 \, \left[ \frac{\ E\left( \frac{4 \tilde{\rho}}{(1+\tilde{\rho})^2} \right)}{\pi (1-\tilde{\rho})}+\frac{\ K\left( \frac{4 \tilde{\rho}}{(1+\tilde{\rho})^2} \right)}{ \pi \  (1+\tilde{\rho})}  \right ]\\  
B_0 &=&  B_z(\rho=0) = \frac{ I }{2R_0} = \frac{ (qe)\ \omega_0 }{4\pi R_0} 
\end{eqnarray} 
where $\tilde{\rho}\equiv \rho/R_0$ and the $K(x)$ and $ E(x)$ are the complete elliptic integral of the first and second kind.  Along the symmetry axis away from the $z=0$ plane, the magnetic field is simply $B_z(z) =  \frac{B_0}{[1+(z/R_0)^2]^{3/2}}$. Clearly one recognizes the existence of magnetic field associated with the swirling charged particle, in line with our general expectation $\mathbf{B} \propto (qe) \bm{\omega}$. It is also easy to see  that the angular momentum of this particle, $L \sim mR_0^2 \omega_0$ is directly proportional to the magnetic flux $\Phi_B \sim B_0 \pi R_0^2 \sim (qe) R_0 \omega_0$ penetrating through the circle, i.e. $L \propto \Phi_B$. 

One can demonstrate the same for a quantum mechanical particle constrained on a 1D circle of radius $R_0$ on x-y plane. In this case the quantum mechanical wave function is simply $\psi = \frac{e^{i  k \phi}}{\sqrt{2\pi}}$ with angular momentum $L = k \hbar$ along $\hat{z}$. The electric current is given by  $I=(qe)(-i\hbar) [\psi^* (\partial_\phi/R_0)\psi - \psi (\partial_\phi/R_0)\psi^*]=\frac{(qe) k \hbar}{\pi R_0} $. Similarly the magnetic field along $\hat{z}$ at the center is given by $B_0= \frac{(qe) k \hbar}{2\pi R_0^2} $, again proportional to the angular momentum, $B_0 \propto L$. So is the magnetic flux, $\Phi_B \propto L$.  \\

\subsection*{Magnetic field of a swirling charged fluid}

We now consider a many-body fluid system that has nonzero vorticity as well as nonzero charge density, as illustrated in Fig.~\ref{fig0} (right). The connection between magnetic field and vorticity in charged fluid could be demonstrated in general. From Maxwell\rq{}s equations we have:
\begin{eqnarray}
\partial_\mu F^{\mu\nu}=J^{\nu}.
\end{eqnarray}
In an ideal fluid with nonzero charge density $n\neq 0$ and fluid field $u^\mu$, the electric current can be expressed as
\begin{eqnarray}
J^\mu= n e u^\mu.
\end{eqnarray}
So we will have:
\begin{eqnarray}
\partial_\mu J_\nu-\partial_\nu J_\mu=ne\omega_{\mu\nu}+(u_\nu\partial_\mu-u_\mu\partial_\nu)ne,
\end{eqnarray}
where $\omega_{\mu\nu}=\partial_\mu u_\nu-\partial_\nu u_\mu$ is the relativistic kinetic vorticity tensor. If the charge density is homogeneous or only very slowly varying, we can keep only the first term in the above and obtain
\begin{eqnarray}
n e \omega^\mu&=& \epsilon^{\mu\nu\rho\sigma}u_\nu\omega_{\rho\sigma}  
= 2 \epsilon^{\mu\nu\rho\sigma}u_\nu\partial_\rho \partial^\lambda F_{\lambda\sigma} 
= \epsilon^{\mu\nu\rho\sigma}u_\nu \Box F_{\rho\sigma}.\label{eq:B-ome-co}
\end{eqnarray}
In the above derivation we have used the relationship $\epsilon^{\mu\nu\rho\sigma}a^\lambda+\epsilon^{\nu\rho\sigma\lambda}a^\mu+\epsilon^{\rho\sigma\lambda\mu}a^\nu+\epsilon^{\sigma\lambda\mu\nu}a^\rho+\epsilon^{\lambda\mu\nu\rho}a^\sigma=0$ and the Gauss-Farady Law $\epsilon^{\mu\nu\rho\sigma}\partial_\nu F_{\rho\sigma}=0$. The above relation clearly demonstrates the direct connection between  vorticity and magnetic field in a charged fluid. This becomes even more transparent for static case in the fluid local rest frame: $ \nabla^2 \bm{B} =  n e \bm{\omega} $ implying a nonzero magnetic field in charged fluid with nonzero vorticity.  While our analysis is based on similar equations as general magnetohydrodynamics (MHD)~\cite{Hernandez:2017mch,Schnack}, we consider  physical systems with nonzero net charge density which is in a regime away from typical ideal MHD analysis (e.g. \cite{Inghirami:2016iru}) and bears different constituent relation for the current. 

\subsection*{A concrete relation for a charged fluid vortex}

To make this connection concrete, let us consider a general fluid vortex structure. In the following analysis we adopt the global lab frame in which the results would be most transparent and more convenient for later application. We describe the vortex with a velocity profile 
\begin{eqnarray}
\bm{v}=v_0\  F\left(\frac{\rho}{R_0}\right) \hat{\mathbf{\phi}} \ .
\end{eqnarray}
The fluid vortex extends along $\hat{z}$ direction with a finite transverse size $R_0$, with the velocity field vanishing for $\rho > R_0$. It should also vanish at the center, i.e. $\bm{v}\to 0$ at $\rho\to 0$. The profile function $F(x)$ is normalized via $\int_0^1 \mathrm{d} x F(x) x = \frac{1}{2}$ so that the averaged velocity of the vortex is $\frac{\int_0^{R_0} \mathrm{d} \rho  \rho \bm{v}}{\int_0^{R_0} \mathrm{d} \rho  \rho} = v_0 \hat{\mathbf{\phi}}$. 
For such a velocity profile, the corresponding vorticity is given by 
\begin{eqnarray}
\bm{\omega} = \frac{v_0}{R_0} \left(\frac{\rho}{R_0}\right)^{-1} \  F\left(\frac{\rho}{R_0}\right) \hat{\mathbf{z}} \ .
\end{eqnarray}
One may  define an average vorticity  $\bm{\bar \omega}$ as
\begin{eqnarray} \label{eq_ave_omega}
\bm{\bar \omega} = \frac{ \int_0^{R_0} \mathrm{d}\rho \rho (n\rho^2) \omega }{ \int_0^{R_0} \mathrm{d}\rho \rho (n\rho^2)} \hat{\mathbf{z}} = \frac{4 v_0}{R_0}\int_0^1\mathrm{d} x F(x) x^2  \, \hat{\mathbf{z}} \ .
\end{eqnarray}
Note in defining the above average, we include a weighing factor $(n\rho^2)$ in a role like moment of inertia which connects angular momentum with vorticity. 

We then solve the corresponding magnetic field from the Maxwell's equation 
$\nabla\times\bm{B}=\bm{J}$
and obtain  
\begin{eqnarray}
\bm{B} =ne v_0 R_0 \int_\frac{\rho}{R_0}^{1}\mathrm{d}x F(x) \ \hat{\mathbf{z}}.
\end{eqnarray}
The magnetic field is the strongest at the center:  
\begin{eqnarray}
B_{max}=n e v_0 R_0 \int_0^{1}\mathrm{d}x F(x).
\end{eqnarray} 
The average magnetic field can be given by 
\begin{eqnarray} \label{eq_ave_B}
\bm{\bar{B}} &=& \frac{\int_0^{R_0} \mathrm{d}\rho \rho \bm{B}(\rho)}{\int_0^{R_0} \mathrm{d}\rho \rho}\hat{\mathbf{z}} 
= 2nev_0 R_0\int_0^{1} \left[\mathrm{d}x x\int_x^1\mathrm{d}x' F(x') \right] \, \hat{\mathbf{z}} 
=  nev_0 R_0\int_0^{1} \mathrm{d} x F(x) x^2  \, \hat{\mathbf{z}} \ .
\end{eqnarray}
where we have done integration by part for the last step.

By comparing Eq.\eqref{eq_ave_omega} and \eqref{eq_ave_B}, we obtain the    key result that connects the  magnetic field with the   vorticity:  
\begin{eqnarray} \label{eq_omega_B}
e\bm{\bar{B}} = \frac{e^2}{4\pi} \  n \ (\pi R_0^2) \ \bm{\bar{\omega}} =  \frac{e^2}{4\pi} \  n \ A \ \bm{\bar{\omega}}
\end{eqnarray} 
where $A=\pi R_0^2$ is the transverse area of the fluid vortex. 
The above relation suggests that there exists an average magnetic field in a charged fluid vortex, which is linearly proportional to the charge density as well as the average fluid vorticity.  This simple relation can be applied as  a new mechanism for generating magnetic field in heavy ion collisions, as we shall discuss next.

\section*{New mechanism for magnetic field in heavy ion collisions}

In heavy ion collisions, there exist nonzero vorticity structures and a nonzero charge density in the created hot fluid. Given the connection between magnetic field and the vorticity in a charged fluid in Eq.\eqref{eq_omega_B},  we propose this as a novel mechanism for the generation of magnetic field in such collisions.  A key factor for this to work, which was not previously studied, is that the considerable net electric charge density (particularly in low beam energy region) would remain in the bulk system during its evolution. In the rest of this Section, we will estimate the magnitude of this new magnetic field for the first time.  We will also show that such magnetic field has considerably long duration as compared with previously known source of magnetic field in these collisions.

\begin{figure*}[htb!]
\begin{center}
\includegraphics[width=16cm]{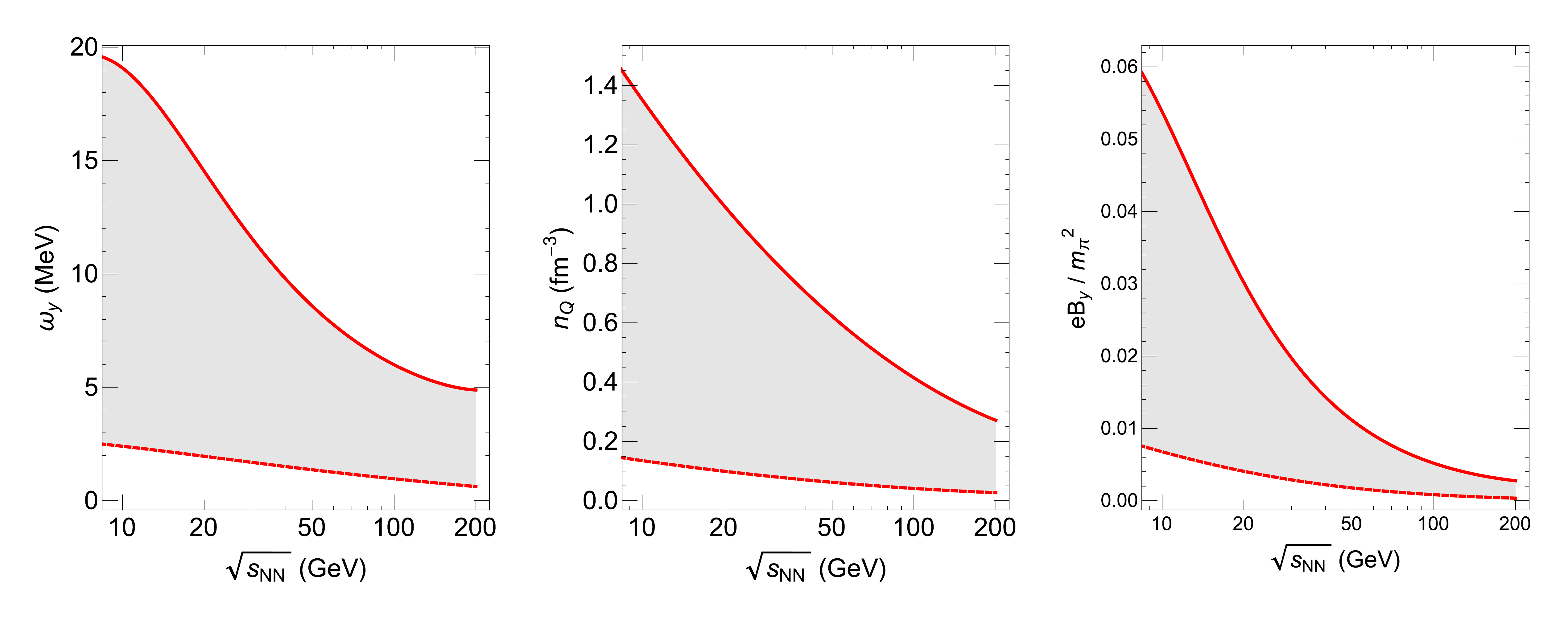}
\caption{\label{fig1} (color online) The vorticity $\omega_y$ (left, in unit of  $\rm MeV$ corresponding to $1.5\times 10^{19} \ sec^{-1}$), charge density $n_Q$ (middle, in unit of  $\rm fm^{-3}=10^{45} \  m^{-3}$) and magnetic field $e\bar{B}$ (right, in unit of  $m_\pi^2$ corresponding to $3.3\times 10^{14}\ \rm Tesla$) as functions of collisional beam energy $\sqrt{s_{NN}}$ (in unit of $\rm GeV=10^9\ \rm eV$), with solid/dashed curves in each panel representing an upper/lower estimates and with the shaded band between them giving an idea of the expected range (see text for details).}
\end{center}
\end{figure*}

The vorticity structures in heavy ion collisions have been computed in various approaches. Let us take   $(20-50)\%$ centrality of AuAu collisions at RHIC in the $(10\sim 200)\rm GeV$ energy region as our example,  which corresponds to the global hyperon polarization measurements by STAR~\cite{STAR:2017ckg}.  One can extract average vorticity $\omega_y$ (along the out-of-plane direction) from AMPT simulations~\cite{Lin:2004en,Lin:2014tya,Shou:2014zsa,Huang:2017pzx,Zhao:2017rpf,Jiang:2016woz,Shi:2017wpk,Li:2017slc} conveniently for a wide beam energy span. Note such vorticity decreases with time in a given collision. We show in Fig.~\ref{fig1} (left) such  average vorticity values (in unit of $\rm MeV$ corresponding to $1.5\times 10^{19} \ sec^{-1}$ ) as a function of beam energy $\sqrt{s}$ for an early time moment $\tau=0.50\ \rm fm/c$ or equivalently $\tau=1.6\times 10^{-24}\ sec$ (solid curve) and a late time moment $\tau=5.0\ \rm fm/c$ or equivalently $\tau=1.6\times 10^{-23}\ sec$ (dashed curve), with the shaded band  giving an idea of the expected range. Clearly the vorticity strongly increases toward low beam energy.

Let us then estimate the charge density $n$ in the fireball. The charge density at late time may be extracted from freeze-out conditions. For example, based on AMPT simulations, one can extract the following parameterization for charge density at freeze-out: $n_{fo} (\sqrt{s_{NN}}) \simeq   0.30 - 0.087 \ln\sqrt{s_{NN}} + 0.0067 (\ln \sqrt{s_{NN}})^2 $ (in unit of $\rm fm^{-3}=10^{45}\  m^{-3}$).  (We note in passing that these estimates are in consistency with chemical freeze-out conditions extracted via thermal models, see e.g.~\cite{Andronic:2017pug,Bzdak:2019pkr}.) The charge density in the fireball  also strongly depends on time due to the fireball expansion and is  significantly larger at earlier time. 
One can verify with explicit AMPT simulations that at the early time the charge density would be about one order of magnitude higher than that at freeze-out time.  We show in Fig.~\ref{fig1} (middle) the charge density values as a function of beam energy $\sqrt{s}$ for an early time moment $\tau=0.5\ \rm fm/c$ or equivalently $\tau=1.6\times 10^{-24}\ sec$ (solid curve) and at freeze-out (dashed curve), with the shaded band between them giving an idea of the expected range. 
The charge density also strongly increases toward low beam energy, due to more  significant stopping effect.

To use Eq.~\eqref{eq_omega_B} for estimating the magnetic field, we still need the area perpendicular to the fluid vortex axis. In our case, that would be the fireball cross-sectional area on the reaction plane (usually labeled $\hat{x}-\hat{z}$ plane).  For AuAu $20-50\%$ collisions the spatial size along the impact parameter ($\hat{x}$ direction) can be reasonably estimated as $R_0\simeq 4\ \rm fm =4\times 10^{-15} \ \rm m$) which grows somewhat toward late time due to transverse expansion. The longitudinal size changes substantially with time due to strong  expansion and also depends on  rapidity window. For higher beam energy collisions, the longitudinal extension is initially small but grows very rapidly. For lower beam energy collisions, the longitudinal extension is not small from the beginning (due to less Lorentz contraction) yet grows less rapidly. In both cases, the relevant longitudinal size would presumably in the plausible range of $(1\sim 10)\rm fm = (1\sim 10)\times 10^{-15}\ \rm m$.  For simplicity we use $A\sim \pi R_0^2 $ with $R_0\sim 4\ \rm fm=4\times 10^{-15}\ m$ as an order-of-magnitude average estimate.  Putting all these together into Eq.\eqref{eq_omega_B}, we thus obtain an estimate for the magnetic field $e \bar{B} $ arising from the charged fluid vortex in heavy ion collisions, as shown in Fig.~\ref{fig1} (right). The solid/dashed curves are obtained from the upper/lower estimates for $\omega_y$ and $n_Q$ (see solid/dashed curves respectively in the left and middle panels), with the shaded band between them giving an idea of the expected range. As one can see, a magnetic field on the order of magnitude $\sim 0.01 m_\pi^2$ (or equivalently $\sim 10^{12}$ Tesla) could be generated through this new mechanism. This magnetic field increases strongly toward lower beam energy. 
In the following we discuss two examples  highly relevant to experimental measurements where this new mechanism may make considerable contributions. 

\begin{figure}[htb!] 
\begin{center}
\includegraphics[width=8.8cm]{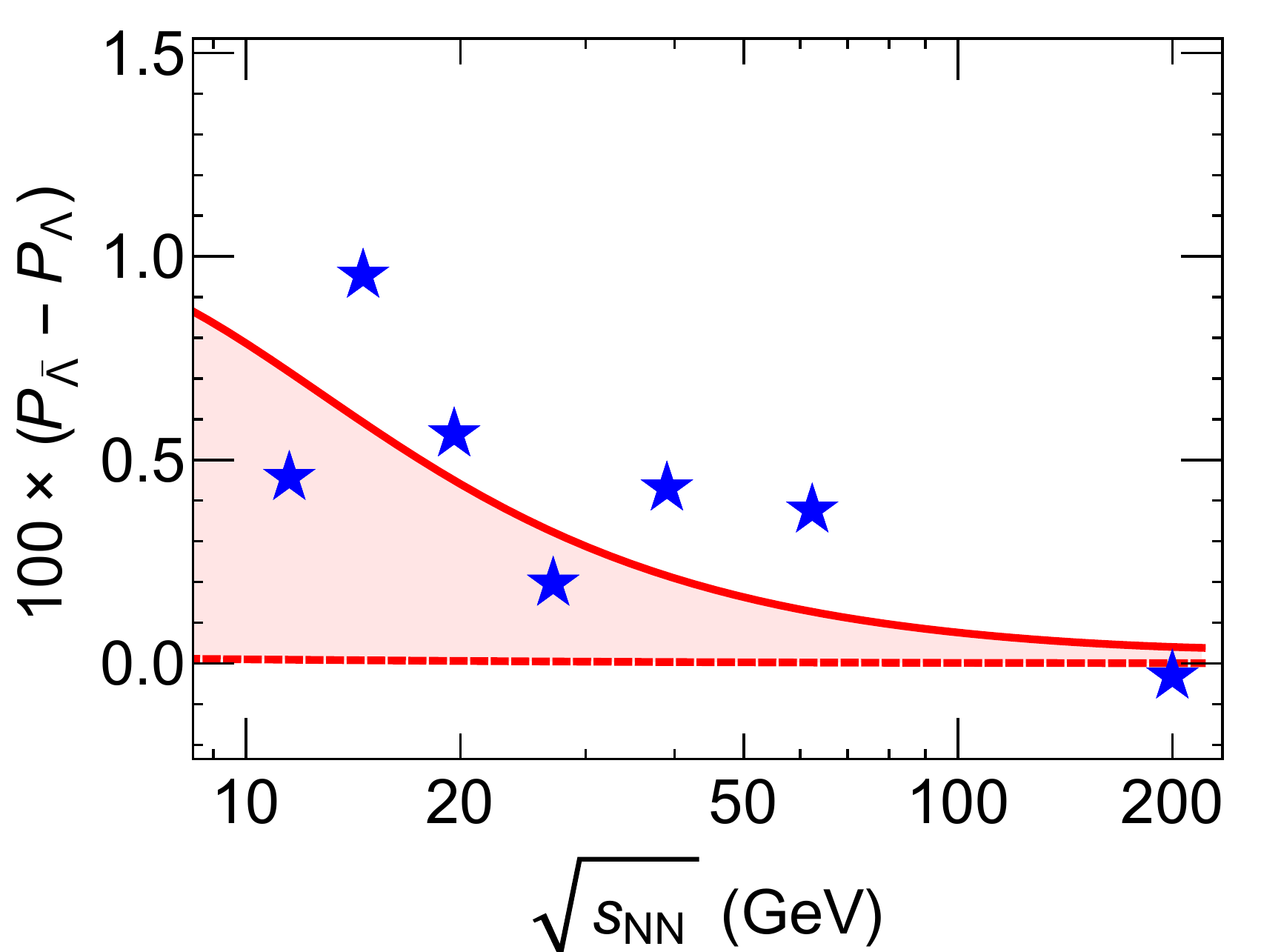}
\caption{\label{fig2} (color online) The induced polarization difference between hyperons and anti-hyperons, $\Delta P = P_{\bar{\Lambda}} - P_\Lambda$ as a function of collisional beam energy $\sqrt{s_{NN}}$ (in unit of $\rm GeV=10^9\ \rm eV$), in comparison with STAR data~\cite{STAR:2017ckg}.   The solid/dashed curves are obtained from the upper/lower estimates for $e\bar{B}$  (see solid/dashed curves respectively in Fig.~\ref{fig1} right panel).}
\end{center}
\end{figure}

\section*{Spin hydrodynamic generation by new magnetic field}

Given the long-lived magnetic field found above, it is natural to examine 
its implication for relevant experimental measurements in heavy ion collisions. As we shall shown in this Section, it turns out to be a novel source of contribution to  the difference in spin hydrodynamic generation for particles and anti-particles. We will also briefly discuss its influence on the CME signal. 

One interesting consequence of such a magnetic field, is its possible contribution to the measured difference in the global polarization of hyperons and anti-hyperons due to their opposite magnetic moments~\cite{Patrignani:2016xqp}. Under the presence of a magnetic field upon freeze-out, one expects: 
\begin{eqnarray}  
\Delta P \equiv P_{\bar\Lambda} - P_{\Lambda} \simeq \frac{2|\mu_\Lambda | \bar{B}}{T_{fo}} 
\end{eqnarray}
where we use $|\mu_\Lambda|  = 0.613 \mu_N=\frac{0.613 \ e}{2 M_N}$ with $M_N=938\rm MeV$~\cite{Patrignani:2016xqp} and $T_{fo} = 155\rm MeV$.  The induced polarization difference $\Delta P$ as a function of beam energy is shown in Fig.~\ref{fig2}, in comparison with STAR data.   Again the solid/dashed curves are obtained from the upper/lower estimates for $e\bar{B}$. 
Despite substantial error bars in current data, the comparison already clearly demonstrates that the proposed new mechanism of magnetic field from charged fluid vortex can induce a considerable difference in the hyperon/anti-hyperon polarizations that could account for a significant portion of the experimental measurements.  This mechanism also leads to a trend in collisional beam energy  that is consistent with the data. Upcoming measurements from the 2nd phase of RHIC beam energy scan program~\cite{Bzdak:2019pkr} would produce much more accurate data to test this mechanism.

Furthermore, such a new magnetic field may bear important impact for anomalous transport effects, such as the Chiral Magnetic Effect (CME) and Chiral Magnetic Wave (CMW), in heavy ion collisions. The signal of these effects would depend upon the time-integrated strength of a magnetic field. Therefore contributions from long-lived magnetic field would be important.   This may be particularly important for relatively lower collisional beam energies such as those available in RHIC Beam Energy Scan experiments. Let us make a simple estimate here. Take the average magnetic field strength to be about $(0.01\sim 0.06) m_\pi^2$ (with $1 m_\pi^2$ corresponding to $3.3\times 10^{14}\ \rm Tesla$) and the lifetime till freeze-out could be estimated as $(5\sim 10)\rm fm/c$ (with $1\rm fm/c= 3.4\times 10^{-24}\ sec$) , the time-integrated strength of the new magnetic field could reach an energy scale in the range of $e\bar{B} \tau \simeq (5\sim 60) \rm MeV $. This new contribution is at similar order of magnitude as  the time-integrated strength of the initial vacuum magnetic field (see e.g. \cite{Muller:2018ibh}). Recent quantitative modeling of CME signals, based on the Anomalous-Viscous Fluid Dynamics (AVFD)~\cite{Jiang:2016wve,Shi:2017cpu}, has also shown that a time-integrated magnetic field of this magnitude can contribute a substantial amount of charge separation signal. Therefore the proposed new mechanism of magnetic field from charged fluid vortex can also influence experimental signals of CME and CMW thus should be taken into account for  modelings of these effects.

\section*{Summary}

We have suggested a novel application of a general link  between rotation and magnetic field in a charged fluid system. This generic connection has been conceptually demonstrated both at single-particle and at multi-particle level. Our analysis has for the first time established this mechanism as a new source for  generating long-lived in-medium magnetic field in heavy ion collisions. Using the relation between magnetic field and vorticity derived in a simple fluid vortex model, estimates have been made for the magnitude of this new magnetic field arising from finite vorticity and net charge density in the colliding systems across a wide span of collisional beam energy. Such a magnetic field is found to rapidly increase toward lower beam energy and has a considerably longer lifespan than previously known source of magnetic field in these collisions. This novel feature has been shown to provide a nontrivial  contribution toward the difference in spin hydrodynamic generation between particles and anti-particles and to account for a significant portion of the previously puzzling experimental measurements. In addition, it is also able to make a considerable contribution to the measurable signal of the Chiral Magnetic Effect. 

We conclude this paper with an outlook into further exploration of this idea. Theoretically, a natural next step would be a quantitative computation of such a new magnetic field by extending a number of current evolution tools for studying magnetic field and vorticity driven effects~\cite{Jiang:2016wve,Shi:2017cpu,Yin:2015fca,Guo:2017jxs}.  
One may also think about ways to experimentally test this idea. The new magnetic field sensitively depends upon the charge density, fluid vorticity and system size. For example, one could contrast different colliding systems like the isobar pairs or compare the AuAu, AuCu, and CuCu colliding systems, which are found to have similar vorticity~\cite{Shi:2017wpk} but different system sizes and charge densities. 
Another possible way is to bin the events  based on their final hadrons' charge asymmetry which is correlated with the charge density in the system and examine how the polarization splitting would vary with the charge asymmetry.

\section*{Acknowledgements }

The authors thank Shuzhe Shi for very helpful discussions. This work is supported in part by the NSFC Grants No. 11735007 and No. 11435004, by the NSF Grant No. No. PHY-1913729 and by the U.S. Department of Energy, Office of Science, Office of Nuclear Physics, within the framework of the Beam Energy Scan Theory (BEST) Topical Collaboration. JL is also grateful to the Institute for Advanced Study of Indiana University for partial support.

\section*{Author contributions}

X.G., J.L. and E.W. contributed equally to this work. All authors reviewed the manuscript.

\section*{Competing interests}

The authors declare no competing interests.

\section*{Data availability}

The computational results generated  and datasets analysed during the current study are available from the corresponding author on reasonable request.

\bibliography{swirl}

\end{document}